# Alloying 2D VSe$_2$ with Pt: from a charge density wave state to a disordered insulator

E. Vélez-Fort[1], P. Mallet[1], H. Boukari[1], A. Marty[2], C. Vergnaud[2], F. Bonell[2], M. Jamet[2] and J-Y. Veuillen[1*]

[1] Univ. Grenoble Alpes, CNRS, Grenoble INP, Institut NEEL, 38000 Grenoble, France

[2] Univ. Grenoble Alpes, CEA, CNRS, Grenoble INP, IRIG-SPINTEC, 38000 Grenoble, France

**\*** : jean-yves.veuillen@neel.cnrs.fr

**Abstract**

We have analyzed by means of scanning tunneling microscopy and spectroscopy the atomic and electronic structure of monolayers of 1T-V$_x$Pt$_{1-x}$Se$_2$ alloys grown by molecular beam epitaxy on epitaxial graphene substrates. We have focused on the composition range (0.1≤x≤0.35) where ferromagnetic behaviour has recently been demonstrated. For low Pt concentration, (x=0.07 and x=0.21), small domains (a few nanometres in diameter) exhibiting the characteristic superstructure of the charge density wave (CDW) state of pristine VSe$_2$ monolayer remain visible on most of the sample surface. Thus alloying preserves the short range order of the CDW phase, although it destroys its long range order. For higher Pt concentration (x≈0.35) a disordered alloy forms. It presents a fully developed gap (a few tens meV in width) at the Fermi level and is thus a disordered insulator. This gap exhibits strong variations at the nanometer scale, reflecting the local fluctuations in the composition. An unexpectedly large interaction of the TMD layer with the graphene substrate sets in for this composition range.

**I. INTRODUCTION**

Since the first reports of ferromagnetism in two-dimensional (2D) materials [1, 2], there has been an intense experimental activity to identify new compounds with high Curie temperature [3-6]. Among them, 1T-VSe2 monolayer is an appealing case due to the prediction of ferromagnetic ground state by ab-initio calculations [7-11] . A pioneering paper has indeed reported the observation of intrinsic high temperature ferromagnetism in this material [12]. However, several later experimental works [13-17] have contested this conclusion. Further reports indicate that the ferromagnetic signal may originate from extrinsic defects [17, 18]. The current understanding is that the absence of ferromagnetic order in defect-free 2D 1T-VSe$_2$ results from a competition with a charge density wave instability, which develops below room temperature for the monolayer [14, 15].

Besides the search for homogeneous phases, one option is to modify, for instance by doping, otherwise non magnetic 2D materials such as transition metal dichacogenides (TMD) to induce magnetic order. Recent achievements are the development of dilute 2D magnetic semiconductors such as V substituted WSe$_2$ [19, 20]. Apart from providing magnetic ions, doping also induces some disorder in the TMD layers; and it is known that a small amount of disorder can reduce significantly the critical temperature for the CDW transition in 2D materials [21]. Based on this result, it has been proposed [22] that doping could depress the temperature at which the CDW order develops in monolayer 1T-VSe$_2$ to sufficiently low values to allow for the establishment of the competitive ferromagnetic state. Further ab-initio calculations have indeed shown that monolayers of ordered 1T-V$_x$Pt$_{1-x}$Se$_2$ alloys could be ferromagnetic for 0.1<x<0.5 [22], even when the lattice deformation associated with the CDW state was taken into account. Measurements on the (presumably disordered) 2D V$_x$Pt$_{1-x}$Se$_2$ alloys confirmed the presence of magnetism at low temperature in MBE grown layers down to the monolayer limit [22]. Specifically, a ferromagnetic order sets-in below 10-30 K in the monolayer phase for x=0.35 [22].

In the aforementioned studies of 2D V$_x$Pt$_{1-x}$Se$_2$ alloys, the detailed characterization of the atomic structure and of the low energy electronic excitations of the material at low temperature was lacking. These points are however relevant for a proper understanding of the nature of the magnetism in the alloys, which are presumably more disordered than the ideal case considered in ab-initio studies. Moreover, the analysis of the low energy electronic structure will indicate whether the material is metallic or not in the temperature range where magnetism develops, which is important for transport studies or possible applications [22]. The structural characterization techniques used in Ref. 22 have already shown that some crystallographic order is preserved upon alloying, with an average lattice parameter following the Vegard's law. Importantly, no intercalated atoms could be detected in the van der Waals gap by means of high resolution electron microscopy (HAADF-STEM), and X-ray diffraction did not reveal any parasitic phases (other than the 1T phase) in the samples [22]. A band structure remains observable in photoemission studies, although the features are broader for alloys than for the pure 1T-VSe$_2$ phase, but little information can be gained on the low energy range (within a few tens meV from the Fermi level).

In this paper, we present a study by scanning tunnelling microscopy (STM) and spectroscopy (STS) performed at 8K of the atomic and electronic structure of monolayers (1L) of 2D V$_x$Pt$_{1-x}$Se$_2$ alloys for the composition range 0.07≤x≤0.35. The samples were grown by MBE on epitaxial graphene layers on SiC, following exactly the same procedure as in Ref. 22 [23]. We focus on the alloys with less than 50% Pt since the structural and magnetic properties of the epitaxial layers have only been thoroughly investigated in this range [22]. Moreover, our preliminary RHEED studies indicate a significant degradation of the crystalline order for x>0.50. For the smallest values of x (x=0.07 and x=0.21), we find that small patches, only a few nanometers in diameter, exhibiting the same reconstruction as the CDW state of pristine 1L 1T-VSe$_2$, still exist in the alloy. A pseudo gap (with typical width 50 meV) is found at the Fermi level in the patches. These nanodomains coexist with disordered regions induced by Pt incorporation. For x≈0.35, the CDW patches have disappeared and the alloy becomes more homogeneous [22], although it presents a granular aspect at the nanometer scale. The electronic structure exhibits a full gap at the Fermi level E$_F$, hereafter called a hard gap. It is essentially symmetric across E$_F$ and its width is of the order of tens of meVs. The 2D V$_x$Pt$_{1-x}$Se$_2$ alloys are thus magnetic insulators for x≈0.35. The gap width shows significant spatial variations at the nanometer scale, which apparently reflect the local distribution of grains. More surprisingly, we find evidences for scattering of the graphene electrons by a sharp potential step at the island edges, which is a priori unexpected for a van de Waals interaction between 1L 1T-V$_x$Pt$_{1-x}$Se$_2$ and graphene. Possible origins for both the gap in the TMD layer and the scattering of graphene electrons are discussed.

## II. RESULTS AND DISCUSSION

### A. Morphology of the films

Since the nominal coverage of the deposited material is lower than one TMD plane, monolayer islands of $V_xPt_{1-x}Se_2$ are readily identified in large scale STM images (see Figure S2 and S3 [23]). We shall focus on this monolayer (1L) phase in the following. To visualize the evolution of the structure of $V_{1-x}Pt_xSe_2$ samples with increasing the Pt content x, medium scale images are shown for the three compositions studied in Fig. 1. These images were taken using essentially the same tunnelling parameters. For the lowest nominal Pt concentration (7%), Fig. 1a, the sample structure consists in small domains with 1D (stripe) reconstruction. Within the islands, the stripes are oriented along three main directions rotated by 120°, as expected for an atomic lattice with hexagonal symmetry as $1T-VSe_2$ or $1T-PtSe_2$. Three domains with such orientations are highlighted by square colour boxes in Fig. 1a, and a numerical zoomed-in image of one domain is shown in the right panel. Within the domains, the perpendicular distance between rows is close to 0.79 nm. An additional order seems to exist along the rows, with a typical spacing of 0.6nm between patterns. The size of the domains is small, with a typical diameter below 5 nm, and their boundaries are decorated by dark spots, which occasionally agglomerate into disordered regions (see the white circles in Fig. 1a). This contrasts with the usually defect-free junctions observed between rotational domains of the 1D superstructure in the CDW state of pure $VSe_2$ [14, 24].

An image for the sample with x=0.21 is shown in Fig. 1b. Its structure is reminiscent from the one of the x=0.07 sample (Fig. 1a), in the sense that "ordered" domains with the 1D stripe reconstruction are still observed, with the same periodicity perpendicular and parallel to the rows. The lateral size (diameter) of the domains remains small (<5nm). The main differences with the x=0.07 sample is that the density of "dark point" defects is significantly larger and, accordingly, the relative amount of "disordered" regions (e.g. white ovals in Fig. 1b) in the islands increase.

From the distance values measured across and along the rows, the structure of the 1D stripe reconstruction for the samples with 7% and 21% Pt is indeed quite similar to the one of the low temperature CDW state of genuine $VSe_2$ [25]. We can thus conjecture that the nanoscale "ordered" domains observed in these samples are actually patches of almost pure $VSe_2$, which present the same CDW superstructure as the genuine phase. This statement is strongly supported by the data presented in Figure 2 that will be be discussed later. One important difference between our Pt doped samples and the pure phase (considering only $VSe_2$ samples grown by MBE on graphitic substrates) is that the size of the "ordered" domains with 1D stripe reconstruction is quite small in our case. Their diameter is smaller than 5 nm, whereas it is (at least) larger than 10 nm for pure $VSe_2$ [17, 24, 25]. Therefore, our data show that Pt incorporation in $VSe_2$ suppresses the long rang order of the CDW state, but not the short range order since small patches of the CDW with nanometer size persist up to x=0.21. Such behavior has already been observed for CDW materials upon increasing the disorder [21, 26, 27], but usually for smaller concentrations of defects. Notice that a recent study [28] also indicates the disappearance of the CDW state for more than 20% doping of $NbSe_2$ with Mo atoms. Another observation is that the short-range CDW-like order already sets-in in domains that are only 2-3 nm in diameter (see also Fig. 2), apparently without the need for a long coherence length [29]. This suggests that a mechanism based on a local lattice deformation or strong coupling, rather than the weak coupling (original Peierls) scenario, is responsible for the development of the reconstruction associated to the CDW state in $VSe_2$ [30, 31]. Possible structures for the lattice distorsion associated with the CDW state in 1L $VSe_2$, and in the related 1L $VS_2$ phase, have been proposed in the literature [14, 15, 17, 32].

Figure 1-c presents an image for the sample with the highest Pt content (x=0.35). The structure of the sample with x=0.33 is quite similar to the one discussed in this paragraph (see Figs. S7 and S11 in Ref. 23). Domains with the 1D stripe reconstruction are no longer present in the monolayer, whatever island is considered. We observe instead a "granular" structure, with "bright grains" whose diameter and separation are in the nanometer range. In detail, the apparent height, width and shape vary slightly from one grain to the other (see also Fig. 3). This granular structure shows up at both positive and negative biases (see Fig. S4 [23]). The distribution of bright grains looks rather uniform on a given island, as already quoted in Ref. 22. However, local fluctuations of the distribution exist. These variations result in spots where the density of grains seems to be larger. Some of those "dense" areas are indicated by light green ovals in Fig. 1c. From Fig. S4 [23], the local density of bright grains remains the same at opposite polarities. In principle, a strictly uniform (ordered) substitution on the V sites would give an average distance between Pt atoms of 0.6 nm for x=0.35. As quoted above, our STM data show variations of the contrast (apparent height) on the TMD monolayer over slightly larger distances, typically a couple of nm's (at least more than 1 nm). This points to a non-uniform distribution of Pt atoms in the material, with local fluctuations of the concentration on the same length scale. We thus conclude that, for nominal Pt content x≥0.33, the domains exhibiting the $VSe_2$-like CDW reconstruction have eventually disappeared and that a disordered $V_{1-x}Pt_xSe_2$ alloy with spatial fluctuations in the V:Pt concentration at the nanometer scale has formed. There is *a priori* no reason to believe that the local atomic structure of this alloy is similar to the one of the disordered regions observed for the x=0.21 sample in Fig. 1b.

## B. Atomic and electronic structure of 1L $V_{1-x}Pt_xSe2$ islands

In this section, we give a more detailed account of the local electronic structure of the islands, as derived from STS studies. Considering the disordered nature of the experimental system, we focus on two points: the structure of the "ordered" domains found in the samples for x=0.07 and x=0.21 and the nature of the gap at the Fermi level for more Pt rich compositions (for x=0.33 and x=0.35).

Fig. 2a displays an atomic resolution image of the x=0.21 sample where small "ordered" domains with two different orientations are found. For this image, the atomic contrast has been enhanced as described in Fig. S5 [23]. This numerical treatment allows visualizing the √3 atomic periodicity along the lines of the stripe 1D reconstruction. In the other direction, we observe either a x2 superstructure perpendicular to the lines or possibly an oblique x√7 one, as shown in the zoomed-in image (right panel) of the green boxed area in Fig. 2a. These two super-periods would lead to a perpendicular distance of 0.69nm or 0.85nm between the lines for VSe$_2$, which is consistent with the value 0.79 nm extracted from the medium resolution images of Fig. 1 [25]. Here, the periodic pattern is limited to 3 nm corresponding to the typical domain size. Nevertheless, the atomic structure displayed in Fig. 2a is almost identical to the one reported for the CDW state of pure VSe$_2$ [12, 14, 16, 17, 24, 25, 33]. This supports our statement that the "ordered" domains with stripe reconstruction in Fig. 1a and 1b are indeed patches of 1L VSe$_2$, which can develop a CDW superstructure despite their small lateral dimensions. Within these domains, we find no evidence for the strong contrast variations due to the V/Pt alloying, which is observed for instance in the x=0.35 sample (see e.g. Fig. 1c and Fig.3). This is another indication of the presence of pure VSe$_2$ patches.

The electronic structure of the green-boxed nano-domain in Fig. 2a was investigated by recording a line of 181 spectra along the 4.50 nm long dashed arrow in Fig. 2a. The line of spectra starts and ends on depressions at the boundary of disordered regions (see raw image in Fig. S5 [23]), but the central part goes across the ordered domain. The resulting conductance map is displayed in Fig. 2b. The signal is rather independent on the position in the central part of the map, which indicates that the electronic structure is homogeneous across the domain, as already found for 1L VSe$_2$ in Ref. [25]. The most salient feature of this map is the strong depletion of the tunneling conductance near the Fermi level (0 bias in Fig. 2b). This characteristic has been reported in all STS studies of the CDW state in genuine 1L VSe$_2$ [12, 16, 25], and it is consistent with the disappearance of at least part of the Fermi surface observed in ARPES studies of this material [13, 16, 24, 33, 34]. This finding is thus in agreement with our proposal for the nature of the ordered nano-domains. The exact nature of depletion of the tunneling conductance, hard gap or pseudogap, remains disputed for 1L VSe$_2$ from the aforementioned STS and ARPES studies. Averaging the spectra over the central part of the domain to enhance the signal to noise ratio, we find (see Fig. 2c) that the conductance indeed goes (almost) to zero right at the Fermi level, but that it reaches small but finite values within 10 meV from E$_F$. This result suggests that there is probably only a pseudogap at the Fermi level in this domain, although we cannot totally exclude the presence of a very small gap (of width ≤10 meV). A similar reduction in the conductance at zero bias corresponding either to a pseudogap or to a very narrow hard gap was consistently found on several domains for the x=0.21 and for x=0.07 (see Fig. S6 [23]) samples. The energy range where the conductance is depressed is about 50 meV. A similar result was reported for pure 1T-VSe2 layers in the CDW phase in Ref. 25.

It has been acknowledged that the development of the CDW phase could allow eluding the magnetic ground state predicted for the bare 1T-VSe$_2$ phase [14, 15]. This competition between CDW and ferromagnetic orders would indeed result in a non-magnetic ground state for 1L VSe$_2$, which agrees with most of the recent experimental reports [13-17]. Qualitatively, this competition can be described using basic arguments. The band structure of 1T-VSe$_2$ exhibit a van Hove singularity at low energy, which leads to a high density of states at the Fermi level [10, 35]. Following Stoner's criterion, a ferromagnetic ground state should appear for the undistorted phase [15, 35]. However, the periodic lattice distortion associated with the CDW phase is expected to lead to a reduction of the density of state at the Fermi level compared to the undistorted phase [15], preventing the development of the ferromagnetic state in 1L VSe$_2$. These results for the genuine 2D material indicate that the "ordered" nano-domains we observe on the 1L islands for the x=0.07 and x=0.21 samples, which we ascribe to patches of 1L VSe$_2$ in the CDW state, are probably non-magnetic. Therefore, the magnetic signal observed for x=0.20 in Ref. 22 should arise from the disordered (or defective) regions indicated in Fig. 1b. Considering that the global (average) magnetization measured in Ref. 22 is larger for x=0.20 than for x=0.35, our data additionnaly suggest that the defective regions we observe in the x=0.21 sample should have a larger magnetic moment per unit surface than the (average) one for the x=0.35 phase. We can not draw such conclusion for the x=0.07 sample since its magnetic structure is unknown: from Ref. 22, it is in between a magnetic phase for x=0.10 and a non magnetic one for x=0.0.

We now turn to the electronic structure of the phase observed for the more Pt-rich samples. We first present in Fig. 3a a picture of the x=0.33 sample with atomic resolution on the 1L TMD island. One can identify the granular structure with the typical length scale in the nanometer range already present in Fig. 1c and Fig. S4. On top of this disordered background, we observe a regular triangular lattice with comparatively weaker contrast (about 10-15 pm), which gives rise to a well-defined set of spots in the Fourier Transformed (FT) image shown in the inset. After calibration on the graphene substrate (see Fig. S7 [23]), we deduce a lattice parameter a≈0.35 nm for the TMD monolayer. This value is intermediate between the in plane lattice parameters of 1T-VSe$_2$ (0.34 nm) and of 1T-PtSe$_2$ (0.38 nm), in agreement with diffraction data [22]. It is usually believed that atomic resolution on TMD layers originates from the top-most Se layer [36-39]. The data of Fig. 3a therefore indicate that a well ordered Se atomic plane persists on top of the V$_{1-x}$Pt$_x$Se$_2$ monolayer, the disorder observed at the nanometer scale being a consequence of the random occupation of the sites in the metal plane by Pt and V atoms.

The voltage dependence of the STM images of the monolayer is illustrated in Fig. 3b to 3d, which have been taken consecutively on the same spot of the x=0.35 sample. We focus there on the low bias values, i. e. V$_s$=±100mV. The reference frame taken at +500mV, Fig. 3b, shows the expected granular structure, with local fluctuations in the density of "bright grains". Examples of areas with high or low grain densities are indicated by light green and black circles respectively. At low sample bias, in Fig. 3c and 3d, the granular structure persists with essentially the same characteristic length scale. However, some grains, and quite often clusters of grains, seem to be "switched off" for V$_s$=±100mV. Bright grains in the low bias images have their counterpart in the image taken at V$_s$=+500mV, and most of the grains which remain visible for V$_s$=+100mV also appear bright for V$_s$=-100mV. Another general remark that can be made is that the bright grains (or clusters of bright grains) which disappear at low bias are often located in low-density areas of the V$_s$=+500mV image (as the one circled in black in Fig. 3b). In the high-density areas (for instance inside the green circle in Fig. 3b) the grains remain bright at low biases. Therefore, it seems that the evolution upon changes in the imaging voltage of the granular structure of the sample consists merely in a disappearance of some bright features at low bias depending on their environment. Notice that, as shown in Fig. S4 [23], the grains which appear bright at V$_s$=+500mV are also bright at V$_s$=-500mV, thus the contrast really depends primarily on the absolute value of the bias. This is an indication for a strong modulation of the low energy tunneling density of states (TDOS) at the nanometer scale in the x=0.33 and x=0.35 samples. It is worth mentioning that the apparent corrugation becomes large at low bias on the V$_{1-x}$Pt$_x$Se$_2$ monolayer, as shown in Fig. 3e, which represents histograms of apparent height for the images in Fig. 3b to 3d. The peak labelled "Gr" close to the origin corresponds to the graphene substrate, and the broad feature labelled "1L TMD" between 0.4nm and 0.8nm gives the distribution of apparent height on the island. The width of this later structure increases dramatically at low bias, and eventually the distribution becomes bimodal reflecting the "switching off" of some bright grains. From current vs. distance curves (not shown), the 0.2 nm increase in the FWHM of the broad peak for $V_s=\pm100$mV would correspond to a decrease by almost two orders of magnitude of the tunneling current in constant height images. This points to very strong variations of the TDOS in this energy range compared to the high bias (±500mV) case.

The spatial fluctuations of the low energy TDOS for the x=0.33 and x=0.35 samples can be analyzed from series of spectra taken along lines crossing the 1L islands. The conductance spectra for such a line are presented in Figure 4. Fig. 4a shows the path followed by the tip (grey dashed arrow) superimposed on images of the area taken at biases of +500mV and +100mV. The line begins on a terrace of the substrate made of Bernal stacked bilayer graphene (BLG) [40]. On the island, the line crosses areas, located close to the island edge and at the end of the line, where the bright grains at +500 mV turn dark at +100 mV. In between, the grains remain bright at both biases. The corresponding conductance map is displayed in Fig. 4b. In this image, the dark blue (yellow) color corresponds to a very low (high) value of the conductance signal. We first notice that, although it is rather homogeneous on the BLG substrate area, the conductance signal on the TMD island shows strong variations on the nanometer length scale. This is especially clear in the low bias range (in the range ±200 mV), where the boundary of the dark blue region can change by as much as 100 mV within less than 2 nm, but it remains true for all biases. These spatial fluctuations in the conductance signal reflect the changes in the TDOS which arise from the disordered V:Pt occupancy of the metal sites (which also give rise to the granular structure quoted above). From the local variations of the contrast in the topographic image of Fig. 4a, we expect, following the discussion of Fig. 3 b-e, that the conductance should remain small at low bias close to the island edge and at the end of the line. The data of Fig. 4a are in agreement with this expectation, showing a strong reduction of the signal in an extended energy range for these locations. We now discuss the existence of a hard gap in the TMD monolayer which corresponds to the dark blue region around zero bias in Fig. 4b.

Conductance spectra on different spots along the line are shown in Fig. 4c and 4d. To increase the signal to noise ratio, especially at low bias, we have averaged d/dV curves on segments of the line where the gap around zero bias (dark blue region) was approximately constant (notice that this procedure would lead to a reduced apparent gap). Horizontal bars with different colors indicate the location of those segments in Fig. 4b. The corresponding spectra, with the same color code, are displayed in Fig. 4c on a linear vertical scale. The curves are shifted vertically for clarity. On the BLG substrate (gray line), we observe a V-shaped pseudogap with a minimum at zero bias, as already reported [41-43]. Although small, the conductance at V=0 mV remains finite. The spectra on the TMD (blue and purple lines) have a U shape at low bias. The width of the minimum where the value of the conductance is strongly depressed amounts to several tens of mV. To demonstrate the presence of a "hard" gap around zero bias on the TMD we plot the logarithm of the dI/dV signal [36] in Fig. 4d. One notices that the conductance indeed reaches a vanishingly small value in a finite bias interval straddling V=0 mV. This establishes the presence of a local gap on the TMD layer, with values ranging from 20 meV (light blue curve) to 60 meV (purple curve). As expected, the gap is larger in the region (purple segment) where the bright grains turn dark at low bias in Fig. 4a (another illustration of this effect is provided in Fig. S9 [23]). The gap is essentially symmetric with respect to zero bias for the three segments shown in Fig. 4c and 4d. From the conductance map of Fig. 4b, as well as from profiles of the current at low bias (see Fig. S8 [23]), we conclude that a hard gap is present all along the line in Fig. 4a. All lines or grids of spectra we have recorded on the $V_{1-x}Pt_xSe_2$ monolayer for x=0.33 and 0.35 gave consistent results. We find a hard gap a few tens of meV wide at all locations on the islands. This gap is rather symmetrical around zero bias and its width exhibits significant variations (by a few tens of meV) at the nanometer length scale. These spatial variations are clearly the spectroscopic counterpart of the granular structure of the sample reported in Fig. 1 and 3.

We now address the possible origin of the hard gap observed at $E_F$ for x=0.33 and x=0.35. It might be a one electron gap (a bandgap), such as the one found for monolayer 1T-PtSe$_2$ [44] or 2H-MX$_2$ with M=Mo, W and X=S, Se [36], with an additional smoothing by the disorder [45]. However, there is no indication for a bandgap in ab-initio calculations [22] of ordered $V_{1-x}Pt_xSe_2$ alloys, neither for x=0.35 nor for neighboring values (x=0.25 and x=0.50). Moreover, even if one assumes that a bandgap exists in this Pt concentration range for a hypothetical ordered alloy, it is difficult to understand how the fluctuations in composition would result in a gap which remains centered at $E_F$. We remind that the composition fluctuates at the local scale (in the nm range) within each sample, and at the global scale between the two samples with x=0.33 and x=0.35. This would lead to strong local doping which shifts the Fermi level to one of the band edges and even to a complete filling of the gap by "impurity states" or by band tails due to the disorder [45]. We thus consider the occurrence of a one-electron gap as unlikely. The granular structure of the sample, which results from the randomness in the occupancy of metal sites as discussed above, is probably a better starting point to grasp the nature of the gap. Disorder alone is not expected to open a gap in an otherwise gapless ordered parent material [45]. Taking electron-electron interactions into account results in a significant depletion of the TDOS at the Fermi level for disordered materials [46]. In their original form, the pioneering models proposed for 2D systems do not lead to a "hard" gap around $E_F$ [47, 48], but to a pseudogap with at most a vanishing TDOS at zero bias only, at variance with our observations (see Fig. 4). Nevertheless, our experimental system is far from the ideal 2D case considered in these models. In our samples, the disordered 2D TMD layer is not freestanding, but it is in direct contact with the conductive graphene plane. This specific configuration allows for a direct transfer of the charges injected by the STM tip in the $V_{1-x}Pt_xSe_2$ to the graphene plane, without the need for a long-range transport of carriers within the TMD layer. The direct path seems to be effective between TMD monolayers and graphene [49, 50]. This perpendicular charge transfer process, together with the granular aspect of the film, suggests an alternative interpretation of the gap in term of a "Coulomb blockade" effect [51-54]. Such mechanism would be consistent with the opening of a symmetric gap at $E_F$. In this scenario, electron tunneling takes place between the graphene and the tip through the TMD layer. Within the TMD layer, disorder would localize the low energy electronic states which are involved in the tunneling process to clusters of "bright grains" of variable sizes. In the spots where the bright grains are closely spaced (dashed circles in Fig. 1c, 3b and Fig. S9 [23]), such states would extend over the whole dense area (this is over a few nm's). Conversely, in spots where the bright grains are further apart, the states would remain localized on the individual grains, i.e. at the nm scale. A smaller Coulomb gap around zero bias should show up in the dense areas because of the larger capacitance related to the increased effective cluster size. Under the localization hypothesis, the (dynamical) "Coulomb blockade" effect would therefore lead to the presence of a gap at $E_F$, with a width varying in space (or locally) according to the local density of bright grains. With this mechanism, the development of a hard gap at zero bias requires that the interfacial resistance between the grains (or clusters of grains) and graphene remains significantly larger than the quantum of resistance [53-55]. A recent report suggests that for molecules held by van der Waals interaction on graphene a double tunnel barrier describes adequately the STS results [56]. Additionally, a significant resistance develops at the interface between large (100 nm²) metallic Pb clusters weakly bound to the graphene substrate [53, 57]. Thus, the existence of a large interfacial resistance between the clusters of grains of $V_{1-x}Pt_xSe_2$ and graphene is *a priori* likely, which would lead to the "Coulomb blockade hard gap" scenario discussed above.

### C. Scattering of graphene electrons at the edges of 1L $V_{1-x}Pt_xSe_2$ islands for x>0.3

To investigate further the Coulomb gap scenario, we have tried to evaluate the interaction between the graphene substrate and the $V_{1-x}Pt_xSe_2$ layer for x≈0.35. How adjacent layers mutually influence their electronic structure is anyway a central issue in designing heterostructures based on 2D materials. Although we are unable to probe the graphene layer buried below the TMD layer by STM, we can estimate the impact of the lateral graphene-TMD interface on the electronic properties of the free (uncovered) material. Figure 5a shows a conductance image of a large area of bilayer graphene (BLG) surrounded by 1L $V_{0.67}Pt_{0.33}Se_2$ islands. The conductance on the free BLG substrate is displayed in orange color, whereas the signal on the TMD layer is saturated (white color). The inset presents the topographic image of the area. The conductance map of Fig. 5a exhibits ripples with a wavelength of approximately 5 nm, which follow the edges of the TMD layer. These ripples reflect the spatial modulations of the local density of states (LDOS) of the BLG [43]. Their observation, and the fact that they are essentially parallel to the island edges, indicate that significant electron scattering takes place at the graphene-TMD boundary. The long wavelength LDOS modulation of Fig. 5a corresponds to intravalley back-scattering in the BLG layer, as reported previously [43]. On the BLG, this process only requires the presence of a potential step at the boundary. This potential can vary slowly on the scale of the unit cell, since intravalley scattering implies only a small wavevector transfer at the scale of the Brillouin zone. However, atomic resolution images (see Fig. 5b) of the BLG substrate taken close to the edge of 1L TMD islands reveal the presence of a √3x√3R(30°) (in short: R3) superstructure extending a few nm's away from the boundary. The R3 superstructure is also observed on SLG, see Fig. S11 [23]. Such patterns involve intervalley scattering in mono and bilayer graphene, which requires a large change in the electronic wavevector [38] and thus an atomically sharp potential. Indeed, the R3 superstructure on graphene or graphite usually shows up around chemisorbed atoms (such as H [58, 59]), vacancies [58, 60] or flake edges [61, 62], which correspond to strong perturbations at the atomic scale. From these considerations, we infer that the potential step at the TMD/graphene lateral interface must vary on the length scale of the graphene unit cell. Evidences for this "atomically sharp" scattering potential has been observed for the two Pt rich samples (see Fig. S10 in Ref. 23 for the x=0.35 sample). It is not present for the more V rich TMD layers for x=0.07 and x=0.21.

The presence of a potential step on the graphene layer at the boundary between the free and the TMD covered regions is of course not surprising. However, the growth of VSe$_2$ [12] or PtSe$_2$ [63] on graphene is supposed to take place by van der Waals epitaxy, which is characterized by the absence of strong covalent bonds between TMD and graphene [64, 65]. Since the van der Waals interaction between the layers is rather long ranged (varying as an inverse power of distance [66]), one would expect a rather smooth scattering potential at the lateral TMD-graphene interface. For the same reason, a potential step of electrostatic origin, due for instance to a charge transfer between graphene and TMD, should vary slowly at the scale of the unit cell. Thus, the basic features of van der Waals epitaxy do not readily explain the existence of the atomically sharp potential step at the lateral graphene/V$_{1-x}$Pt$_x$Se$_2$ interface. Our data thus suggest that for x≈0.35 the interaction between graphene and V$_{1-x}$Pt$_x$Se$_2$ might be stronger than expected for a van der Waals heterostructure. The disorder might play a role in this effect, although its characteristic length scale (at least one nanometer from STM images) is again significantly larger than the graphene lattice parameter. A significant coupling between the TMD and the underlying graphene layer may seem inconsistent with the large interfacial resistance required to open a hard gap in the Coulomb blockade mechanism invoked above. Notice however that from our measurements we cannot discriminate between a strong interaction taking place below the whole TMD flakes or limited only to the island edges. Moreover, the analysis of the electronic structure of the free graphene layer does not provide a quantitative estimate of the coupling strength below the TMD islands. Clearly, this point would deserve further investigations.

### III. CONCLUSION

We have analyzed the atomic and electronic structure of monolayers of V$_{1-x}$Pt$_x$Se$_2$ alloy deposited on epitaxial graphene in the composition range 0.07<x<0.35 by means of low temperature STM. For the lowest Pt content (x=0.07 and x=0.21), we find that the superstructure characteristic of the charge density wave (CDW) state of the pure VSe2 monolayers is preserved in small domains with a diameter of only a few nanometers. Thus, alloying destroys the long-range order of the CDW phase, but preserves the short-range order. For higher Pt content x≈0.35 a more homogeneous alloy forms, although it presents a nanometer scale granular structure. This disorder arises from the random substitution of Pt atoms on the metal sites, which induces sizable fluctuations of the STM contrast and of the electronic properties at the nanometer scale. Noticeably, it presents a hard gap a few tens of meV wide, which seems to be correlated with the local fluctuations of composition. We tentatively ascribe this gap to charging effects, although other mechanisms cannot be ruled out. Finally, our measurements have revealed the presence of a strong electron scattering on the uncovered graphene layers at the boundaries with the V$_{1-x}$Pt$_x$Se$_2$ islands for x≈0.35. Such scattering is unexpected from the basic characteristics of van der Waals epitaxy, and it suggests a significant graphene/TMD coupling at least on the island edges. Anyway, such scattering may be detrimental for devices where charge transport across the lateral TMD-graphene interfaces plays a central role, and the presence of this effect should be evaluated carefully.

**ACKNOWLEDGMENTS**

The authors acknowledge the financial support from the ANR project MAGICVALLEY (ANR-18-CE24-0007).

**FIGURE CAPTIONS**

**FIG. 1.** Morphology of the $V_{1-x}Pt_xSe_2$ monolayer for different compositions. For each value of x, the left panel displays a 40x25 nm² image acquired with the sample bias -500mV. The images were processed to enhance the contrast on the $V_{1-x}Pt_xSe_2$ monolayer, so that the graphene substrate and the second TMD layer regions appear as black and white areas respectively. The right panel is a numerical zoom with size 5x5 nm² on the area marked by the purple square in the left panel. a) x=0.07. The purple, blue and green squares indicate domains of the 1D stripe reconstruction with three different orientations. The white ovals indicate regions which appear as disordered due to the agglomeration of dark point defects. b) x=0.21. The purple square and the white ovals highlight the same features as in a). c) x=0.35. The green ovals indicate areas where the « bright grains » are more closely packed (named "dense areas" in the text).

**FIG. 2.** Atomic and electronic structure of the sample with x=0.21. a) Left panel: 16.0x8.0 nm² image taken with sample bias Vs=-100 mV and tunneling current It=1.0 nA. The image has been processed to enhance the atomic resolution as detailed in figure S5 in section SI3. The image of the right panel is a numerical zoom with a size 3.0x3.0 nm² on the area indicated by the green square in the left panel. The unit cells of the √3x2 and √3x√7 superstructure of the striped 1D reconstruction are shown as blue and black dotted diamonds respectively. b) Conductance map built from a series of 181 dI/dV spectra taken along the line indicated by the dashed blue arrow in a). The setpoint for the spectra is Vs=-500mV, It=500pA. The dark blue color corresponds to very low conductance. c) Average of 40 spectra in the box limited by the dashed green lines in panel b), where the conductance signal depends only weakly on the position. Inset: zoom in the low bias region (+/-30mV) with a linear conductance scale. Sample temperature: 8K

**FIG. 3.** Atomic structure of the Pt rich samples (x=0.33 and x=0.35). a) Atomic resolution image taken on the sample with x=0.33. The hexagonal lattice with lattice constant a=0.35 nm is superimposed on the granular structure. Image size: 8.0x8.0 nm², sample bias Vs=+500mV, tunneling current It=50pA. The Fourier transform in the inset reveals the spots of the reciprocal lattice. b) to d) Variable bias images of the same area with size 35.1x9.5 nm² for the sample with x=0.35. The contrast is enhanced on the 1L TMD, so that graphene and 2L TMD zones appear dark and white respectively. The sample bias Vs is given on each image, the blue scale bar represents 5.0 nm. Green (black) circles indicate spots with high (low) density of bright grains in the high bias image of b). These areas turn bright (dark) respectively in the low bias images c) and d). e) Height histogram on the area of images b) to d) from full scale images (without contrast enhancement). The peak around zero apparent height comes from the graphene substrate, the structure between 0.4 nm and 0.8 nm corresponds to the monolayer $V_{0.65}Pt_{0.35}Se_2$. This later peak broadens on the low apparent height side at low biases (±0.1V).

**FIG. 4.** Electronic structure of the Pt rich samples for x=0.33. a) Images of the edge of a monolayer TMD flake laying on a BLG substrate at high (Vs=+500 mV, left panel) and low (+100mV, right panel) sample bias. Image size: 7.3x20.0 nm². b) Conductance map built from 164 spectra taken along the grey dashed line in a). The setpoint for the spectra is Vs=+500 mV and It=200 pA. The bias range in which a very low conductance (dark blue color) persists is much more extended on the TMD than on the BLG. c) and d) Average conductance spectra for selected regions along the spectroscopic line shown in b). The color of the spectra corresponds to the regions indicated by horizontal bars on the zero bias line in b): grey is for BLG, and shades of blue for different spots on the TMD. c) Spectra displayed using a linear vertical scale (the curves are offset for clarity). A broad U shaped minimum with variable width is observed at all locations on the TMD. d) Same spectra displayed using a logarithmic scale. This representation shows that the conductance is vanishingly small in a finite bias range around zero on the TMD, which corresponds to a hard gap. The noise level is evaluated on the curve with the widest gap. Sample temperature: 8 K.

**FIG. 5.** Interaction between the TMD and graphene for the Pt rich sample (x=0.33). a) Conductance signal (taken simultaneously with the topographic image shown in the inset) on a free area of BLG substrate surrounded by 1L TMD islands. Image size: 150x55 nm² (scale bar: 20 nm), sample bias: +100 mV. The ripples on the BLG area are a standing wave pattern whose wavelength is close to 5 nm. b) Atomic resolution (topographic) image on another free BLG area close to the boundary with monolayer TMD islands. Image size: 18.0x9.0 nm² (scale bar: 2.0 nm), sample bias: +100mV. The pink rectangular box and the blue oval indicate regions which display the (1x1) and the √3x√3R(30°) (R3 in brief) superstructure of BLG respectively. c) Fourier transform of b). The spots of the (1x1) and of the R3 superstructure of graphene are highlighted by color circles.

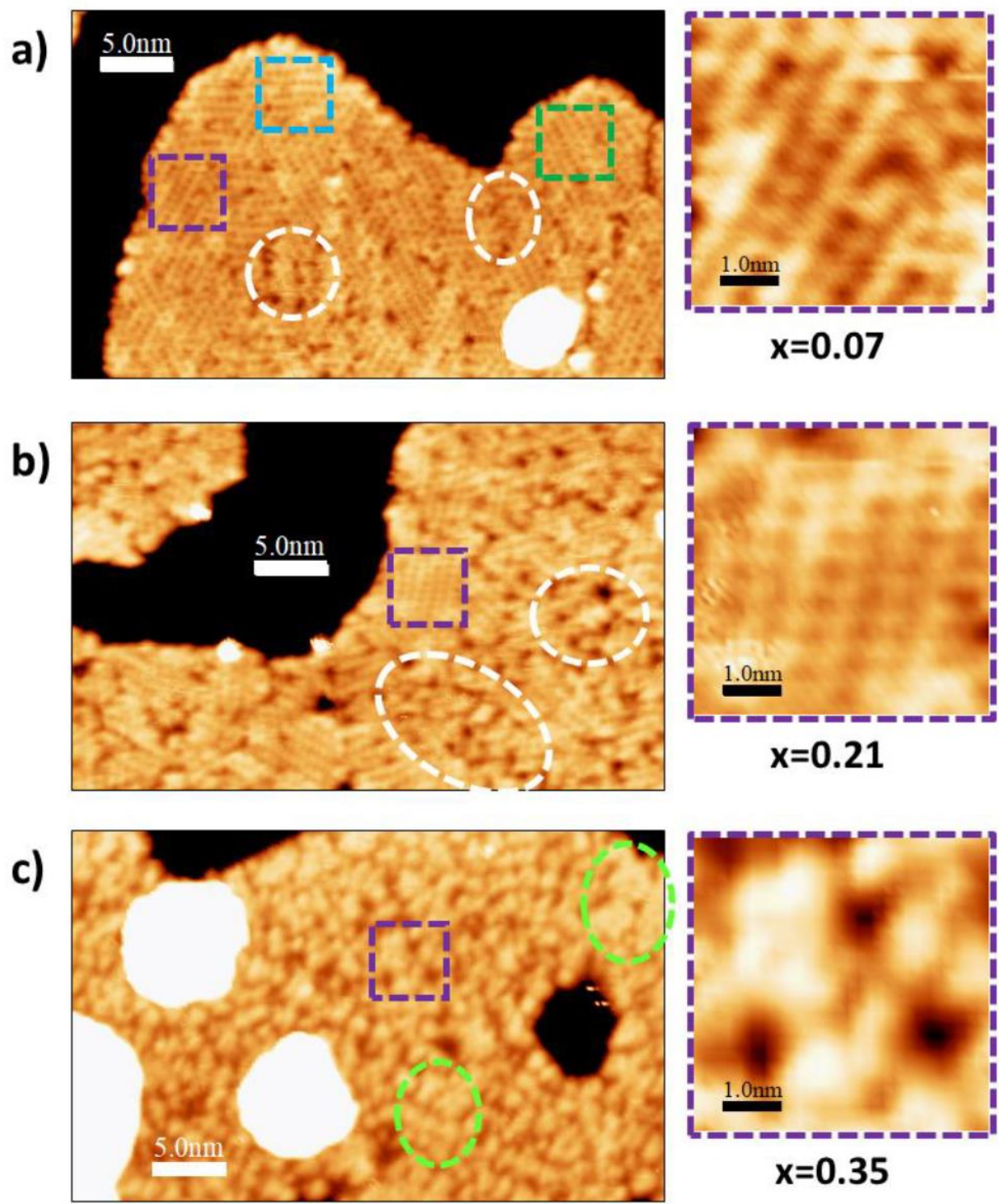

**FIGURE 1**

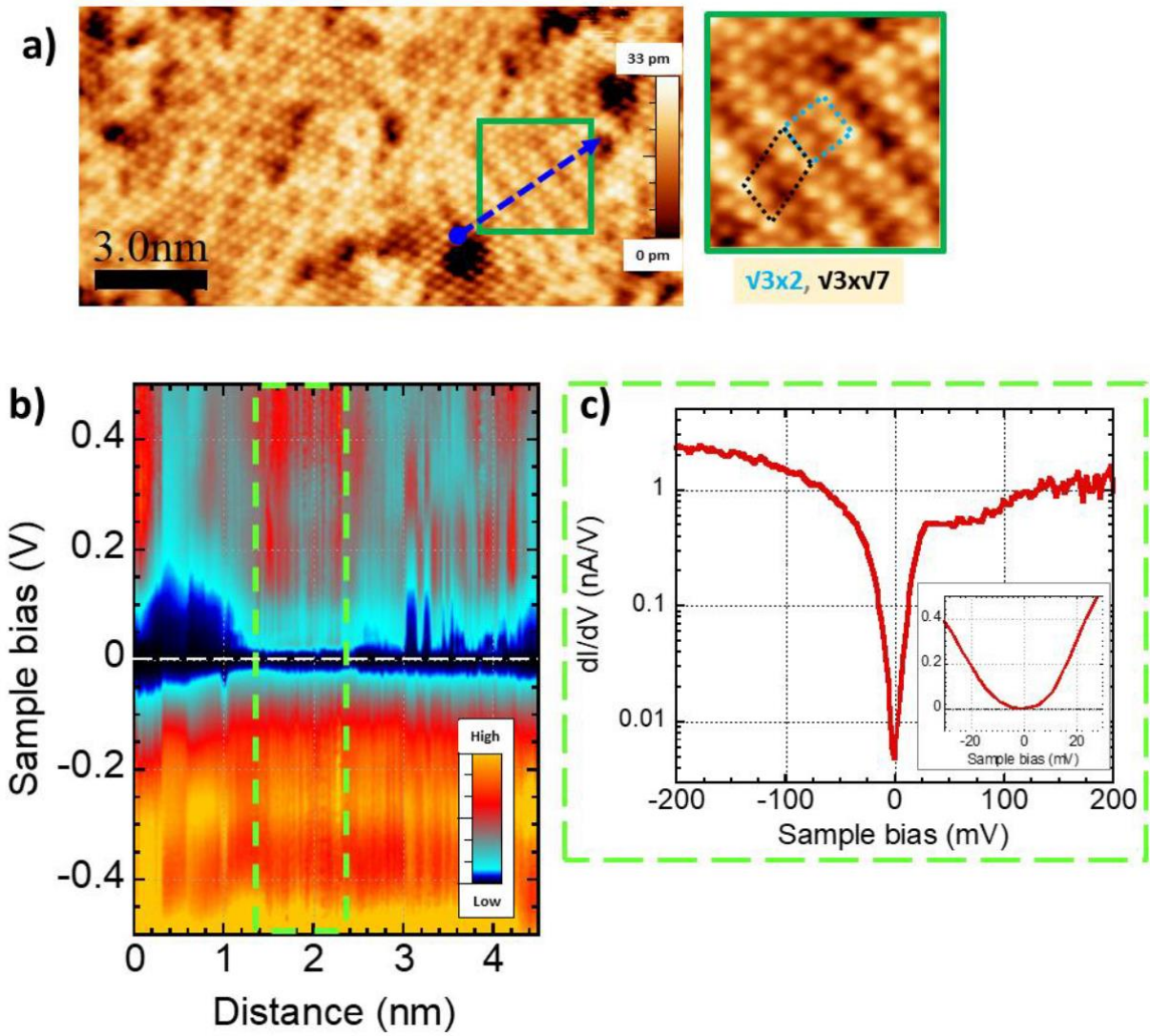

**FIGURE 2**

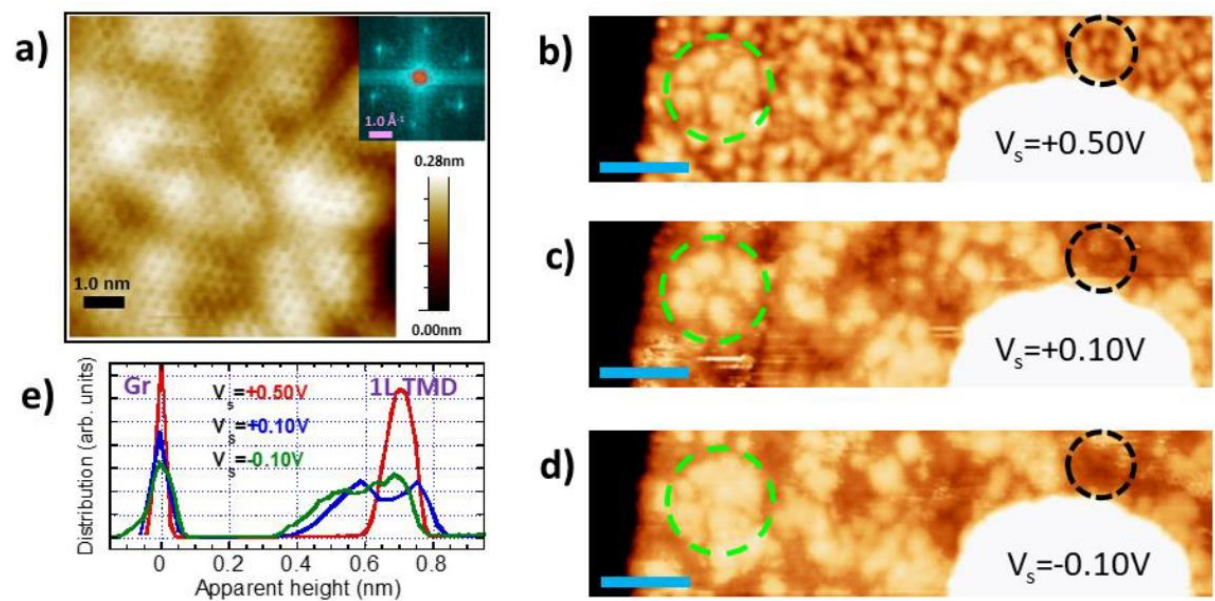

**FIGURE 3**

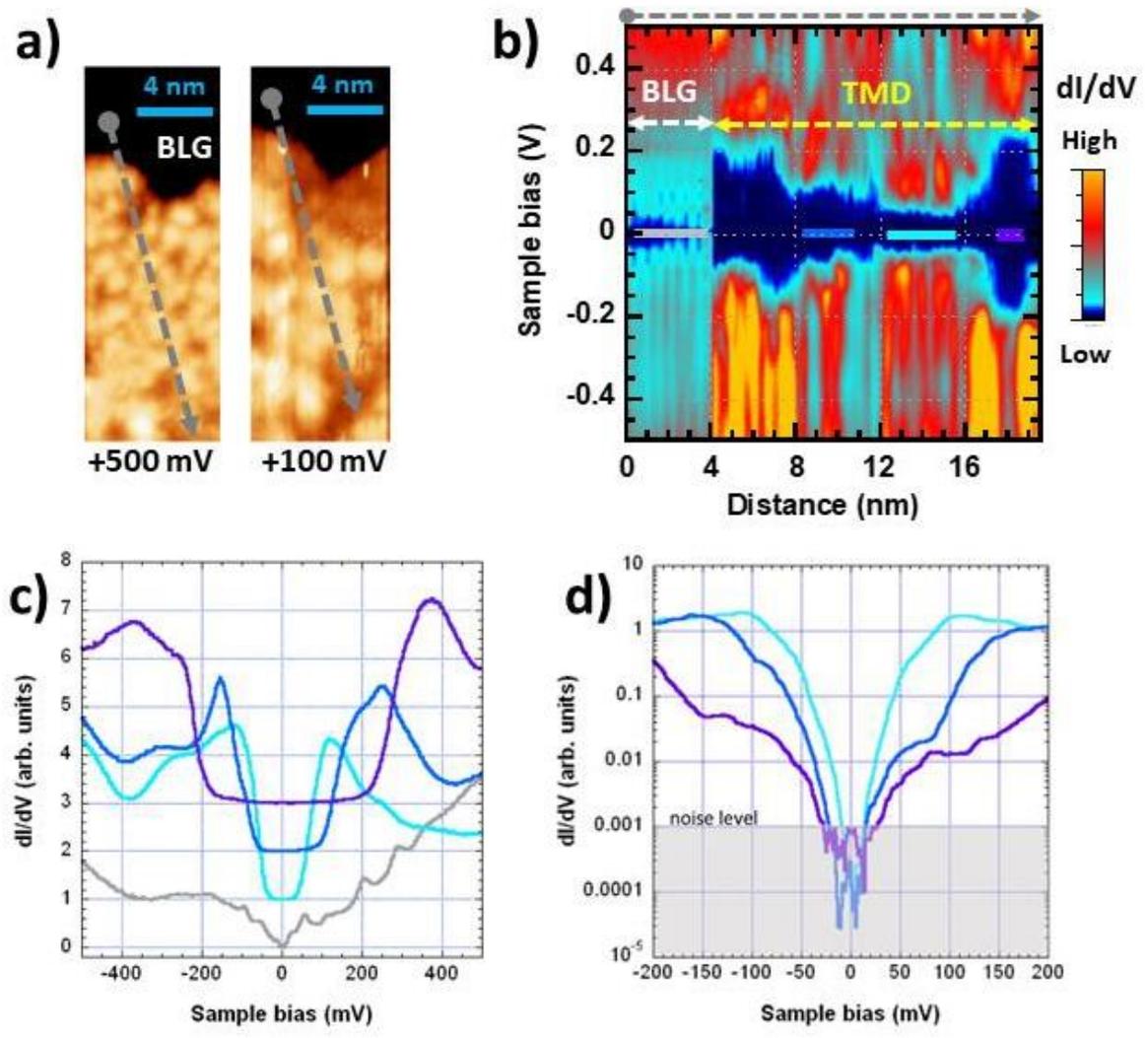

**FIGURE 4**

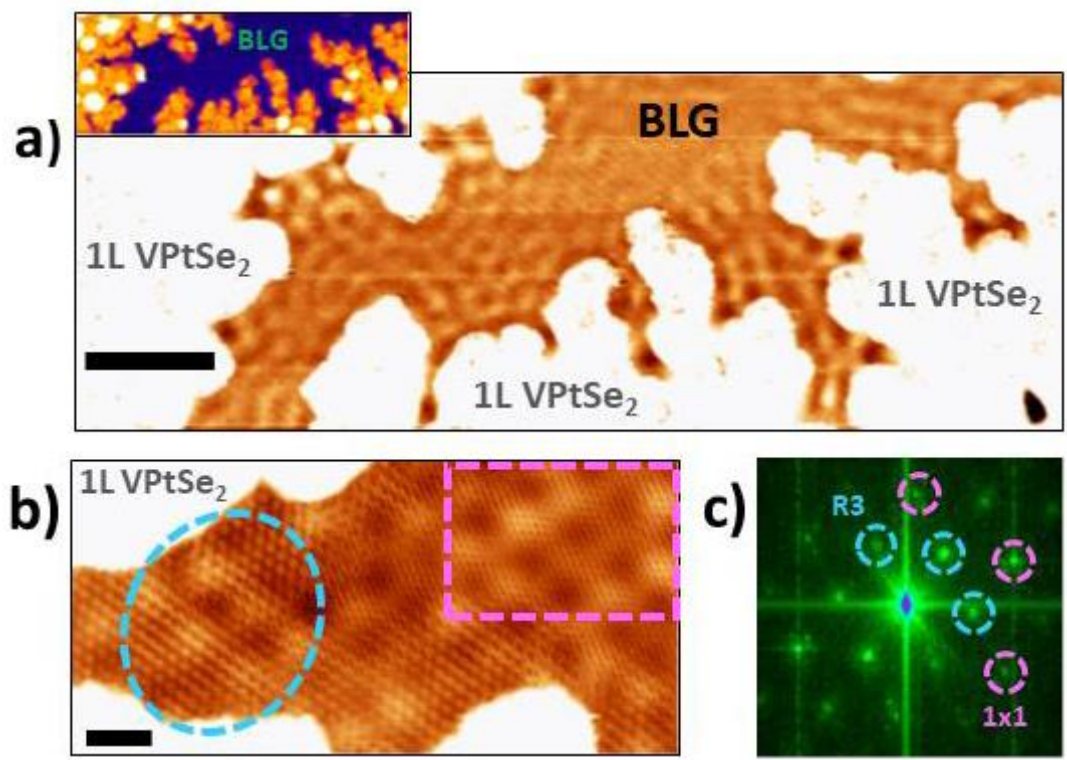

**FIGURE 5**